\documentclass[aps,prd,onecolumn,showpacs,groupedaddress]{revtex4-2}
\usepackage{graphicx}
\usepackage{dcolumn}
\usepackage{bm}
\usepackage{color}
\usepackage{longtable}
\usepackage{supertabular}
\usepackage[T1]{fontenc}
\usepackage{epstopdf}
\usepackage{amsmath}
\usepackage{amssymb}
\usepackage{setspace}
\usepackage{multirow}
\usepackage{booktabs}
\usepackage[section]{placeins}
\usepackage{mathrsfs}
\usepackage{hyperref}

\makeatletter

\newcommand{\Rmnum}[1]{\expandafter\@slowromancap\romannumeral #1@} 
\newcommand{\bq}{\begin{equation}}
\newcommand{\eq}{\end{equation}}
\newcommand{\bqn}{\begin{eqnarray}}
\newcommand{\eqn}{\end{eqnarray}}
\newcommand{\nb}{\nonumber}

\makeatother

\begin{document}

\title{On statistical fluctuations in collective flows}

\author{Wei-Liang Qian$^{2,1,3}$}\email[E-mail: ]{wlqian@usp.br}
\author{Kai Lin$^{4, 2}$}
\author{Chong Ye$^{1,3}$}
\author{Jin Li$^{5}$}
\author{Yu Pan$^{6}$}
\author{Rui-Hong Yue$^{1}$}

\affiliation{$^{1}$ Center for Gravitation and Cosmology, College of Physical Science and Technology, Yangzhou University, Yangzhou 225009, China}
\affiliation{$^{2}$ Escola de Engenharia de Lorena, Universidade de S\~ao Paulo, 12602-810, Lorena, SP, Brazil}
\affiliation{$^{3}$ Faculdade de Engenharia de Guaratinguet\'a, Universidade Estadual Paulista, 12516-410, Guaratinguet\'a, SP, Brazil}
\affiliation{$^{4}$ Hubei Subsurface Multi-scale Imaging Key Laboratory, Institute of Geophysics and Geomatics, China University of Geosciences, 430074, Wuhan, Hubei, China}
\affiliation{$^{5}$ College of Physics, Chongqing University, 401331, Chongqing, China}
\affiliation{$^{6}$ College of Science, Chongqing University of Posts and Telecommunications, 400065, Chongqing, China}

\begin{abstract}
In relativistic heavy-ion collisions, event-by-event fluctuations are known to have non-trivial implications.
Even though the probability distribution is geometrically isotropic for the initial conditions, the anisotropic $\varepsilon_n$ still differs from zero owing to the statistical fluctuations in the energy profile.
On the other hand, the flow harmonics extracted from the hadron spectrum using the multi-particle correlators are inevitably subjected to non-vanishing variance due to the finite number of hadrons emitted in individual events.
As one aims to extract information on the fluctuations in the initial conditions via flow harmonics and their fluctuations, finite multiplicity may play a role in interfering with such an effort.
In this study, we explore the properties and impacts of such fluctuations in the initial and final states, which both notably appear to be statistical ones originating from the finite number of quanta of the underlying system.
We elaborate on the properties of the initial-state eccentricities for the smooth and event-by-event fluctuating initial conditions and their distinct impacts on the resulting flow harmonics.
Numerical simulations are performed.
The possible implications of the present study are also addressed.
\end{abstract}

\date{Jan. 19th, 2022}

\maketitle
\newpage

\section{Introduction}\label{section1}

Strong collective motion features the largely thermalized matter produced in relativistic heavy-ion collisions.
The primary characteristics of the system are indicated by the flow harmonics extracted from the azimuthal correlations between the final state particles.
Relativistic hydrodynamics constitutes one of the most promising theoretical frameworks to describe the temporal evolution of the underlying strongly coupled quark-gluon plasma~\cite{hydro-review-04, hydro-review-05, hydro-review-06, sph-review-01, sph-review-02, hydro-review-07, hydro-review-08, hydro-review-09, hydro-review-10}.
As an effective theory at the long-wavelength limit, such an approach models the system in terms of a continuum.
It plays a vital role in understanding the relationship between the empirical observables and the initial conditions dictated by the underlying microscopic approach. 
The latter furnishes the initial conditions as the input of the hydrodynamic model, primarily expressed in terms of the density distribution, which is subject to event-by-event fluctuations.
Even though specific details of the initial state might not survive the temporal evolution, it is generally understood~\cite{hydro-vn-07, hydro-vn-08, hydro-v3-01, hydro-v3-02, sph-eos-02, sph-cfo-01, sph-corr-ev-10, sph-corr-03, sph-corr-04, sph-corr-05, sph-corr-ev-06, sph-corr-ev-08, sph-corr-09} that the collective flow carries crucial information on the hot and dense system created in the heavy-ion collisions.
These relevant harmonics $v_n$ are defined as the Fourier coefficients of the one-particle distribution function in azimuthal angle $\varphi$~\cite{event-plane-method-1}
\begin{eqnarray}
f(\varphi)=\frac{1}{2\pi}\left[1+\sum_{n=1}2v_{n}\cos{n(\varphi-\Psi_n)}\right] \equiv f_1(\varphi),
\label{oneParDis}
\end{eqnarray}
where the reference orientation $\Psi_n$ for a given order $n$ is known as the event plane, 
In particular, the elliptic flow $v_2$ is mostly attributed to the geometric almond shape of the initial system~\cite{hydro-vn-08}.
The triangular flow $v_3$ is due to the event-by-event fluctuations of the initial conditions~\cite{hydro-v3-01}.
Many studies have been devoted to investigating the relationship between the initial geometric anisotropy and the final-state flow harmonics~\cite{hydro-v3-02, hydro-vn-34, hydro-vn-42, sph-corr-ev-04, sph-vn-04, hydro-corr-04, hydro-corr-05, sph-corr-ev-06, sph-corr-ev-08, sph-corr-09}.

Regarding the analysis of the initial conditions for hydrodynamics, the anisotropies are measured by the complex eccentricities $\epsilon_n$~\cite{hydro-v3-02}
\begin{eqnarray}
\epsilon_n=\frac{\langle z^n\rangle}{\langle |z|^n\rangle}\equiv \frac{\int dx dy  \rho(z)z^n}{\int dxdy \rho(z)|z|^n} ,
\label{epsDef}
\end{eqnarray}
where $z=x+iy$. 
The average $\langle\cdots\rangle$ is performed for a given event, defined on the transverse plane $x-y$ weighted by the energy density $\rho$.
Also, one assumes that the center of the density coincides with the origin.
Moreover, the cumulant $\kappa_n$ has turned out to be a helpful tool to extract further the {\it connected} part of the anisotropies~\cite{hydro-v3-02, hydro-corr-ph-38, hydro-vn-51}, following the exponential formula in combinatorial mathematics.
Specifically, the cumulant can be derived using the logarithm of the generating functional of moments.
The literature often assumes that $v_n$ is primarily proportional to $|\epsilon_n|$ for individual events.
In other words, the probability of $|\epsilon_n|$ is, up to a proper rescaling, that of the $v_n$ distribution.
Due to fluctuations, even though the probability distribution $\rho(z)$ for the initial conditions is geometrically isotropic, the resulting eccentricities do not vanish for $n\ne 0$.
A notable example is the identical point-source model subject to an isotropic Gaussian average density profile, as proposed by Ollitrault~\cite{hydro-vn-08}.
There, it was shown~\cite{hydro-vn-42, hydro-vn-51} that the cumulants $\epsilon_2\{2\}$ and $\epsilon_2\{4\}$ do not vanish as long as the number of partonic constitutes remains finite.

On the side of the collective flow, to extract the harmonics $v_{n}$ from the experimental data, one needs to adopt a specific estimation scheme.
The conventional event plane method~\cite{event-plane-method-1, event-plane-method-2, event-plane-method-3} aims to estimate the event planes $\Psi_n$ in Eq.~\eqref{oneParDis} and subsequently the flow harmonics $v_n$.
The approach is somewhat plagued by the fact that the reaction plane fluctuates on an event-by-event basis~\cite{hydro-v3-01} and cannot be directly measured experimentally.
One can detour the difficulty by forming particle pairs, where the event planes are canceled.
In this regard, many other approaches have been primarily based on particle correlations.
In particular, the mathematical formalism can be concisely expressed in the generating function~\cite{hydro-corr-ph-03, hydro-corr-ph-04}.
This class of approaches consists of the multi-particle cumulant~\cite{hydro-corr-ph-03, hydro-corr-ph-04, hydro-corr-ph-10, hydro-corr-ph-23}, Lee-Yang zeros~\cite{hydro-corr-LY-zeros-01, hydro-corr-LY-zeros-02, hydro-corr-LY-zeros-03}, and symmetric cumulants~\cite{hydro-corr-ph-36}, among others recent generalizations~\cite{hydro-corr-ph-27, hydro-corr-ph-42}.

In practice, even though the accessible number of events is significant, for an individual event, the particle multiplicity is finite.
When averaging over the events, the variance of the extracted harmonic coefficients remains finite even though the particles emanated independently according to a well-defined one-particle distribution function.
Such a degree of statistical uncertainty will present itself as flow fluctuations.
In practice, it might become indistinguishable from those caused by the event-by-event fluctuations in the initial conditions discussed above. 
The latter is understood to reflect the physics of the underlying microscopic model, different from its counterpart of statistical origin.

The present study is mainly motivated to explore the above aspect in flow fluctuations.
We examine how the initial geometric fluctuations and finite multiplicity give rise to the variance in the measure flow harmonics and multi-particle correlators.
The remainder of the present paper is organized as follows. 
In the next section, we briefly review the identical point-source model and derive the fluctuations of eccentricities.
In Sec.~\ref{section3}, we study the statistical fluctuations in the estimated harmonic coefficients due to the finite multiplicity.
These two aspects are simultaneously taken into consideration in Sec.~\ref{section4}, where the main results of the current paper are presented.
The last section is devoted to further discussions regarding the implication of flow analysis in relativistic heavy-ion collisions and the concluding remarks.  

\section{Eccentricities and its fluctuations}\label{section2}

In this section, we address the eccentricity fluctuations owing to the finite number of participants, which eventually form ``nuggets'' in the initial energy distribution.
This characteristic is largely demonstrated in many well-known event generators, such as MC Glauber~\cite{glauber-review-1}, CGC MC-KLN~\cite{cgc-mckln-01, cgc-mckln-02, cgc-mckln-13, cgc-mckln-14}, NeXUS~\cite{nexus-1, nexus-rept}, EPOS~\cite{epos-1, epos-2, epos-3}, among others.
As shown in the specific energy profiles of the generated initial conditions (see, for instance, Fig.~2 of Ref.~\cite{cgc-ip-glasma-03}, Fig.~1 of Ref.~\cite{sph-review-02}, and Fig.~2 of Ref.~\cite{epos-9}), the event-by-event fluctuating initial conditions typically feature a granular energy profile.
The primary idea is that the density fluctuations even on top of an isotropic density distribution will break the symmetry of the average distribution.
For the present purpose, one may consider that the initial conditions are composed of sources owing to partonic binary collisions, as in the context of the Glauber model.
For mathematical convenience, these sources and their fluctuations are generated by a given identical form and are spatially uncorrelated.
As the number of the sources $N$ becomes significant, one may utilize the multivariate central limit theorem (CLT), where the number of variables is governed by the relevant degree of freedom of the initial distribution.
Subsequently, the mathematics simplifies and the derived feature is shown to be largely universal~\cite{hydro-vn-08}.
This is because the CLT concerns the eventual convergence to a multivariate normal distribution of an average among the sampled independently distributed random variables.
The latter, in practice, can be taken as the coordinates on the transverse plane.
To derive the distribution of eccentricity $\epsilon_2$, one may follow Ref.~\cite{hydro-vn-06, hydro-vn-08} by performing a change of variables into new variables that include eccentricity.
When feasible, the resultant eccentricity distribution is obtained by integrating out the remaining variables. 

In what follows, we consider a simplified scenario to illustrate eccentricities generated by fluctuations.
To be specific, one employs the identical point-source model~\cite{hydro-vn-08, hydro-vn-51}, where the sources are point-like, and the probability distribution satisfies an average Gaussian profile
\begin{eqnarray}
    p(z_i)= \frac{1}{\pi R_0^2} \exp\left(-\frac{|z_i|^2}{R_0^2}\right) , \label{eveICprofile}
\end{eqnarray}
where the location of source $i$ is denoted by $z_i$, where $i=1,\cdots, N$.
In this case, the resultant probability distribution of $|\epsilon_2|$ can be derived analytically, which is found to be~\cite{hydro-vn-08}
\begin{eqnarray}
    P\left(|\epsilon_2|\right)=(N-2)|\epsilon_2| \left(1-|\epsilon_2|^2\right)^{\frac{N}{2}-2}, \label{pEps2}
\end{eqnarray}
where $N$ is the number of point sources.
By integrating Eq.~\eqref{pEps2}, it is straightforward to show that
\begin{eqnarray}
    \mathrm{E}\left[ |\epsilon_2| \right] &=& \frac{\sqrt{\pi}\Gamma\left(\frac{N}{2}\right)}{2\Gamma\left(\frac{N+1}{2}\right)}, \nb\\
    \mathrm{E}\left[ |\epsilon_2|^2 \right] &=& \frac{2}{N}, \nb\\
    \mathrm{E}\left[ |\epsilon_2|^4 \right] &=& \frac{8}{2 N + N^2}, \label{EEps2}
\end{eqnarray}
where for a positive interger $M$, the $\Gamma$ function simplies $\Gamma(M)=(M-1)!$.
The variance of $|\epsilon_2|$ are readily given by Eq.~\eqref{EEps2} as
\begin{eqnarray}
    \mathrm{Var}\left[ |\epsilon_2| \right] = \frac{4}{N^2}- \frac{\pi\Gamma\left(\frac{N}{2}\right)^2}{4\Gamma\left(\frac{N+1}{2}\right)^2} . \label{EVarEps2}
\end{eqnarray}
The expected value and variance discussed here refer to the average between different events.

Moreover, the cumulants of the eccentricity are found to be~\cite{hydro-vn-08}
\begin{eqnarray}
    \epsilon_2\{2\} &=& \sqrt{\mathrm{E}\left[ \epsilon_2|^2\right]}=\sqrt{\frac{2}{N}}, \nb\\
    \epsilon_2\{4\} &=& \left(2\mathrm{E}\left[ |\epsilon_2|^2\right]^2-\mathrm{E}\left[ |\epsilon_2|^4\right]\right)^{1/4} = \left(\frac{16}{N^2(N+2)}\right)^{1/4} . \label{cumulantsEps2}
\end{eqnarray}

The above results indicate that the eccentricities do not vanish due to the finite participants.
In the continuum limit, however, eccentricities do not persist, as it is apparent that $\epsilon$, its variance, and the cumulants vanish in the limit $N\to \infty$.
In this regard, it is worth noting that eccentricities generated by initial state geometric fluctuations do not necessarily be implemented in a discrete fashion.
Even at the limit $N\to \infty$, continuous fluctuations in the initial conditions also give rise to non-vanishing eccentricities.
In particular, analyses have been carried out using an expansion in $1/N$ and the magnitudes of the fluctuations~\cite{hydro-corr-ph-07, hydro-corr-ph-08, hydro-corr-ph-11}.

\section{Variance of the flow harmonics}\label{section3}

Although the flow harmonics is formally defined by Eq.~\eqref{oneParDis}, in practice, one needs to utilize a {\it statistical estimator} to extract its value from the empirical data.
The latter inevitably leads to a degree of uncertainty due to the finite multiplicity.

In literature, most methods to extract the flow harmonics are based on particle correlations, and the cornerstone of such an approach is based on the following $k$-particle correlation~\cite{hydro-corr-ph-23}
\begin{eqnarray}
    \langle k\rangle_{n_1,\cdots,n_k}\equiv \langle e^{i(n_1\varphi_1 + \cdots + n_k\varphi_k)} \rangle ,  \label{nkCorr}
\end{eqnarray}
where $\varphi_i$ is the azimuthal angle of the $i$th particle, the average is taken for all distinct tuples of particles.
To focus on the flow harmonics $v_n$, one usually chooses a specific set of $(n_1, \cdots, n_k)$ so that $\sum_{j=1}^k n_j=0$.
For instance, in the case of two-particle correlation $k=2$, one often considers $n_1 = -n_2 = n$.

For a realistic event composed of finite multiplicity, the average $\langle\cdots\rangle$ on the l.h.s. of Eq.~\eqref{nkCorr} is carried out as a summation for all distinct tuples.
Also, any auto-correlation must be properly removed.
One therefore introduces
\begin{eqnarray}
\widehat{k}_{\{n_1, n_2, \cdots n_k\}}\equiv
\frac{\sum\limits_{k-\mathrm{tuples}} w_{1}w_{2}\cdots w_{k} e^{i(n_{1}\varphi_1+n_{2}\varphi_2+\cdots+n_{k}\varphi_k)}}{\sum\limits_{k-\mathrm{tuples}} w_{1}w_{2}\cdots w_{k}}
    = \frac{\mathrm{N}_{\langle k \rangle_{n_{1},n_{2}\cdots n_{k}}}}{\mathrm{D}_{\langle k \rangle_{n_{1},n_{2}\cdots n_{k}}}} .
\label{MultiK_dis}
\end{eqnarray}
where the formalism has been further generalized to include weight $w_k$~\cite{hydro-corr-ph-36}.

In practice, the numerator and denominator of Eq.~\eqref{MultiK_dis} can be expressed by employing the Q-vectors~\cite{hydro-vn-07}, given by
\begin{eqnarray}
    Q_{n,p}\equiv \sum_{j=1}^{M}w_{j}^{p}e^{in\varphi_{j}} ,
\label{qVec}
\end{eqnarray}
where $p$ is an exponent that can be chosen conveniently to simply the resultant expressions.
As an example, for $k=2$, one has
\begin{eqnarray}
     \mathrm{N}_{\langle 2\rangle _{n_1, n_2}} &=& Q_{n_1,1}Q_{n_2,1}-Q_{n_1+n_2,2} , \label{eqN8}\\
     \mathrm{D}_{\langle 2\rangle _{n_1, n_2}} &=& Q_{0,1}^{2}-Q_{0,2} ,\label{eqD8}
\label{eqQvector2p}
\end{eqnarray}

To be more specific, let us consider a pair of particles that are emitted independently for an individual event according to the one-particle distribution function.
Therefore, $k=2$, and one considers $w_1 = w_2 =1$, $n_1 = -n_2 = 2$. 
By denoting the azimuthal angles of the pair by $\varphi_1$ and $\varphi_2$, it is readily verified that the expected value of $e^{i(\varphi_1-\varphi_2)}$ gives 
\begin{eqnarray}
    \langle 2 \rangle_{2,-2} \equiv \langle e^{i2(\varphi_{1}-\varphi_{2})}\rangle 
    = \langle\cos{2(\varphi_{1}-\varphi_{2})}\rangle = v_2^2, \label{v22inCumu}
\end{eqnarray}
where the average 
\begin{eqnarray}
\langle\cdots\rangle=\int d\varphi_1d\varphi_2 \cdots f(\varphi_1, \varphi_2) \nb
\end{eqnarray}
is evaluated by integrating out the azimuthal angles using the joint probability 
\begin{eqnarray}
    f(\varphi_1, \varphi_2)=f_1(\varphi_1) f_1(\varphi_2), \label{findf1}
\end{eqnarray}
where $f_1$ is defined by Eq.~\eqref{oneParDis}.
In other words, one has considered the statistical limit of infinity multiplicity. 

On the other hand, in practice, for an event of a finite number of multiplicity $M$, one may estimate $v_2^2$ by the following form
\begin{eqnarray}
    \widehat{2}_{\{2, -2\}}\equiv \widehat{v_2^2}= \frac{1}{M(M-1)}\sum_{\substack{i \ne j}}\cos{2(\varphi_{i}-\varphi_{j})} ,
\label{Estimatorv2c2}
\end{eqnarray}
where the summation enumerates all distinct $M(M-1)$ ordered pairs. 
It is worth point out that Eq.~\eqref{Estimatorv2c2} is not identical to but a discrete version of Eq.~\eqref{v22inCumu}.

In this regard, Eq.~\eqref{Estimatorv2c2} is a statistical estimator~\cite{book-statistics-Rice} of the physical quantity $\theta = v_2^2$, denoted as $\hat{\theta}$, for an individual event.
Mathematically, the quality of an estimator is measured in terms of its expected value and variance regarding the event average.
If the underlying distribution, namely, Eq.~\eqref{findf1}, is known beforehand, we have
\begin{eqnarray}
    \mathrm{E}\left[\widehat{2}_{\{2, -2\}}\right] &=& v_2^2 ,\label{muv22}\\ 
    \mathrm{Var}\left[\widehat{2}_{\{2, -2\}}\right] &=& \frac{1+v_{2}^2}{M(M-1)}+2\frac{M-2}{M(M-1)}v_{2}^{2}(1+v_{4}) + \frac{(M-2)(M-3)}{M(M-1)}v_{2}^{4}-v_{2}^{4}. \label{sigma2v22}
\end{eqnarray}
Again, we note the difference between the event average denoted by $\mathrm{E}\left[ \cdots\right]$ and that for a given event between particle tuples denoted by $\langle\cdots\rangle$ by assuming infinite multiplicity.

It is apparent that Eq.~\eqref{muv22} indicates that the estimator Eq.~\eqref{Estimatorv2c2} is unbiased, while Eq.~\eqref{sigma2v22} gives an uncertainty of the estimation, between different events of finite multiplicity $M$.
Even though the value of $v_2$ is well-defined in Eq.~\eqref{oneParDis}, again, multiplicity inevitably gives rise to a finite variance.
The latter will be mixed up with and, to a certain degree, indistinguishable from those due to the initial state fluctuations.
These flow fluctuations occur on an event-by-event basis and do not vanish unless, for instance, one generates an infinite number of particles in a hydrodynamic simulation.

Also, the physical nature of the variance shown in Eq.~\eqref{sigma2v22} is rather different from that given by Eq.~\eqref{EVarEps2}.
The former is understood to be governed by the underlying microscopic model and whose expected value does not vanish at a significant number of events.
The latter is purely statistical. 
As long as the estimator is unbiased, the expected value tends to approach the true value when the number of events becomes significant, while the variance persists on an event-by-event basis.
Furthermore, depending on the quality of the estimator, different flow estimation schemes might give different variances.
As the flow evaluation scheme must be applied to real events with finite multiplicity, it is inevitably subject to some statistical uncertainty.

Before closing this section, we elaborate on another example of flow fluctuations due to finite multiplicity associated with particle correlation estimator.
We consider $k=3$, $w_1 = w_2 = w_3 =1$, $n_1 = n_2 =2$, $n_3=-4$, and also assume $\Psi_2=\Psi_4=\Psi$ for simplicity.
By explicit integration, one finds
\begin{eqnarray}
     \mathrm{E}\left[\widehat{3}_{\{2, 2, -4\}}\right] 
      &=& v_2^2 v_4 ,\label{p3N}\\
      \mathrm{Var}\left[\widehat{3}_{\{2, 2, -4\}}\right] 
      &=& \frac{1}{M(M-1)(M-2)}\left\{\left[4v_2^2v_4+2v_4^2v_8\right]\right.\nb\\
      &+& (M-3)\left[4v_2^4+8v_2^2v_4^2+4v_2^2v_4v_8+2v_4^3\right] \nb\\
      &+& (M-3)(M-4)\left[4v_2^4v_4+v_2^4v_8+4v_2^2v_4^3\right] \nb\\
      &+& \left.(M-3)(M-4)(M-5)v_2^4v_4^2\right\} - v_2^4v_4^2 .\label{eqVar3b}
\end{eqnarray}
Although a bit tedious, the evaluations of the above expressions are straightforward.
All possible combinations involve picking out particle pairs from two ordered tuples, which consist, respectively, of three distinct particles.
One enumerates all different possibilities where one, two, or three particles from the two tuples coincide.
Owning to finite multiplicity $M$, the results given in Eqs.~\eqref{sigma2v22} and~\eqref{eqVar3b} demonstrate the fact that in general the estimators $\widehat{k}_{\{n_1, n_2, \cdots n_k\}}$ are subject to finite uncertainty.
In reality, the event planes $\Psi_2$ and $\Psi_4$ do not coincide precisely.
Subsequently, the three-particle correlator $\widehat{3}_{\{2, 2, -4\}}$ actually carries the information on the event plane correlation~\cite{hydro-corr-ph-27, hydro-corr-05}.

\section{Flow variance due to statistical and initial geometric fluctuations}\label{section4}

In this section, we turn to discuss the scenario where both factors discussed in the two preceding sections are taken into consideration.
As discussed above, in a realistic event, the flow fluctuations, demonstrated in terms of the variance of flow harmonics, carry the information on both initial-state geometrical and final-state statistical fluctuations.
We first show the effect numerically using Monte Carlo simulations and then present the analytical results on the resultant flow fluctuations. 

\begin{figure}[ht]
\begin{tabular}{cc}
\vspace{-26pt}
\begin{minipage}{250pt}
\centerline{\includegraphics[width=1.2\textwidth,height=1.0\textwidth]{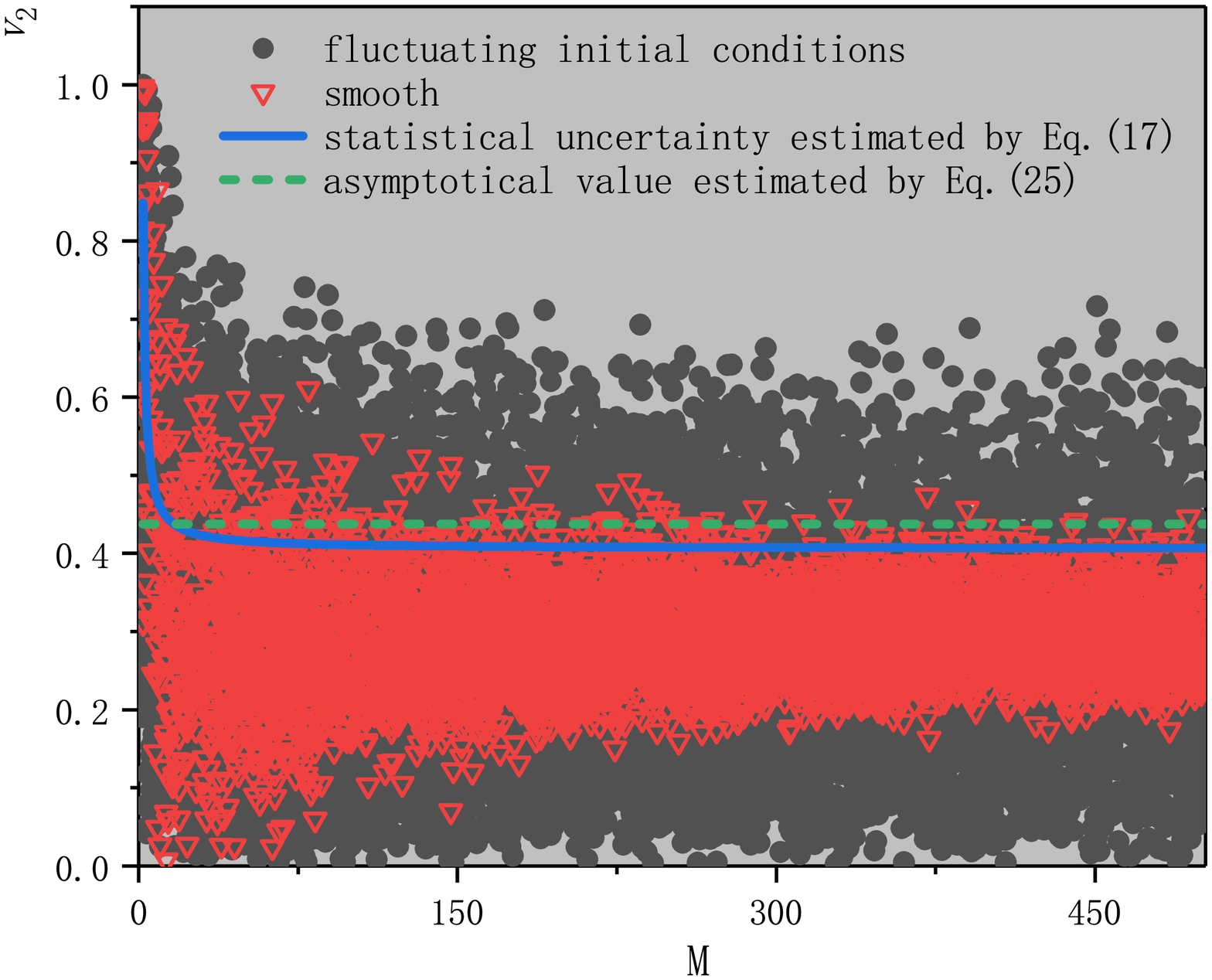}}
\end{minipage}
&
\vspace{22pt}
\begin{minipage}{250pt}
\centerline{\includegraphics[width=1.2\textwidth,height=1.0\textwidth]{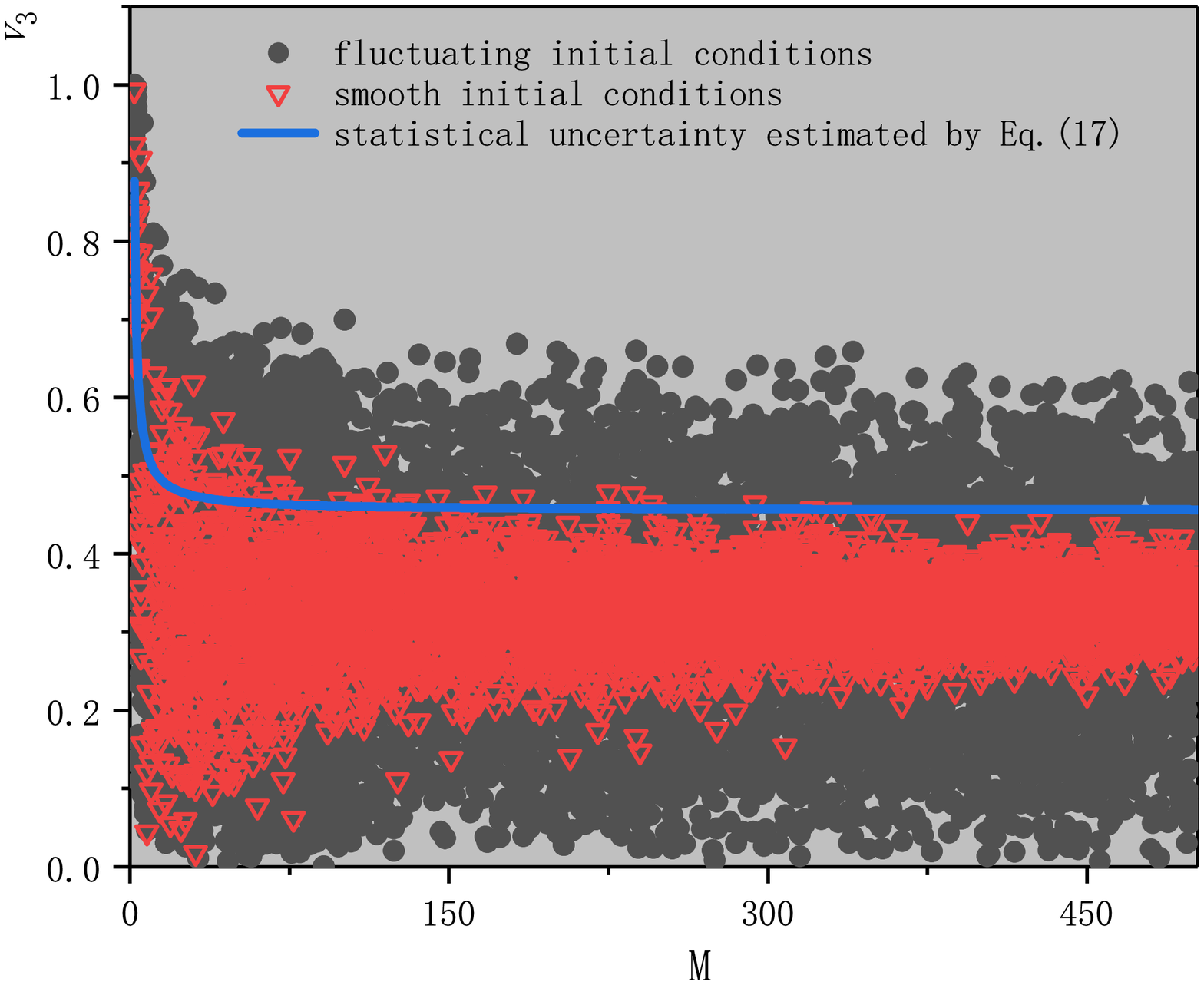}}
\end{minipage}
\end{tabular}
\renewcommand{\figurename}{Fig.}
\caption{The extracted elliptic and triangular flow coefficients for the events of smooth and fluctuating initial conditions, as a function of multiplicity per event $M$ for $N=10$.
The values of $v_2$ are estimated using $\sqrt{\widehat{v_n^2}}$ given by Eq.~\eqref{Estimatorvnc2}.
The hollow red triangles represent the results obtained by the smooth initial condition, and the filled black circles are those by the identical point-source model.
The solid blue curve estimates the statistical uncertainties using Eq.~\eqref{sigma2v22}.
The dashed green curve shows the asymptotical value of the square root of the variance given by Eq.~\eqref{sigma2v22lim} due to event-by-event fluctuations.}
\label{fig_v2_fluc}
\end{figure}

In Fig.~\ref{fig_v2_fluc}, we present the obtained elliptic and triangular flow coefficients for the generated events of smooth and fluctuating initial conditions.
For fluctuating initial conditions, they are prepared according to Eq.~\eqref{eveICprofile}, where one assumes the number of nuggets $N=10$.
Then the eccentricities $\epsilon_n$ are extracted using the definition Eq.~\eqref{epsDef}, where the recentering correction is considered. Namely, one replaces $z$ with $z'=z-z_0$, where
\begin{eqnarray}
z_0 = \frac{\int dx dy z\rho(z)}{\int dx dy \rho(z)}.
\end{eqnarray}
This leads to $\epsilon_1=0$ for all the events.
One can readily verify the numerical implementation by considering the case $N=2$, which, by the definition Eq.~\eqref{epsDef}, one has $\epsilon_2=1$, free of the event-by-event fluctuations.
For simplicity, we adopt the usual assumption that the genuine flow harmonics are proportional to the eccentricities, 
\begin{eqnarray}
v_n = \mathcal{C}_\mathrm{v}\epsilon_n .\label{vnCen}
\end{eqnarray}
For illustrative purposes, it is further assumed that the proportional constant $\mathcal{C}_\mathrm{v} = 1$.
These flow coefficients, in turn, determine the one-particle distribution given by Eq.~\eqref{oneParDis}.
In the present calculation, the flow harmonics in Eq.~\eqref{oneParDis} are truncated at $n=5$.
We employ a Monte Carlo procedure to generate the hardons to simulate the realistic events with a finite multiplicity $M$.
Subsequently, the values of $v_n$ are estimated by the square root of $\widehat{v_n^2}$, which is a straightforward generalization of Eq.~\eqref{Estimatorv2c2}.
To be specific,
\begin{eqnarray}
    \widehat{2}_{\{n, -n\}}\equiv \widehat{v_n^2}= \frac{1}{M(M-1)}\sum_{\substack{i \ne j}}\cos{n(\varphi_{i}-\varphi_{j})} ,
\label{Estimatorvnc2}
\end{eqnarray}
where in our calculations, $n=2, 3$.
As a result, the estimated flow harmonics bear the statistical uncertainty due to finite multiplicity in the particle emission according to Eq.~\eqref{oneParDis}.
The results are presented as a scatter plot of estimated flow harmonics as a function of the event multiplicity, $v_n$ vs. $M$.

For the events with smooth initial conditions, the calculations are rather similar, except that the fluctuations in eccentricities are frozen.
One will always take their mean values on the r.h.s. of Eq.~\eqref{vnCen}.
Again, even though the initial Gaussian distribution is isotropic, the average eccentricities do not vanish.

From Fig.~\ref{fig_v2_fluc}, it is observed that the flow harmonics are subjected to fluctuations, which decrease with increasing event multiplicity.
As expected, the geometrical fluctuations in the initial state give rise to additional uncertainties of flow harmonics on top of those statistical ones.
This is demonstrated as the scatters from events of smooth initial conditions mostly sit in a narrower region on top of those generated by fluctuating initial conditions.
Moreover, as the event multiplicity increases, such a difference becomes more significant.
According to Eq.~\eqref{sigma2v22}, for the events of smooth initial conditions, the variance of flow harmonics decreases with increasing event multiplicity and vanishes at the $M\to \infty$.
On the other hand, for the events generated by fluctuating initial conditions, although the variance of flow harmonics decreases with event multiplicity, they approach a constant value, as further discussed and evaluated below.
The above results are confirmed by the quantitative values presented in Tab.~\ref{tabeveN10}, where one estimates the elliptic and triangular flows and their variance.
In order to show the robustness of our conclusion, we also carried out calculations using a more significant number $N=50$.
The latter results are presented in Tab.~\ref{tabeveN50}.
It is observed that the main feature persists: finite multiplicity gives rise to additional flow variance on top of those due to event-by-event initial conditions.

\begin{figure}[ht]
\begin{tabular}{cc}
\vspace{-26pt}
\begin{minipage}{250pt}
\centerline{\includegraphics[width=1.2\textwidth,height=1.0\textwidth]{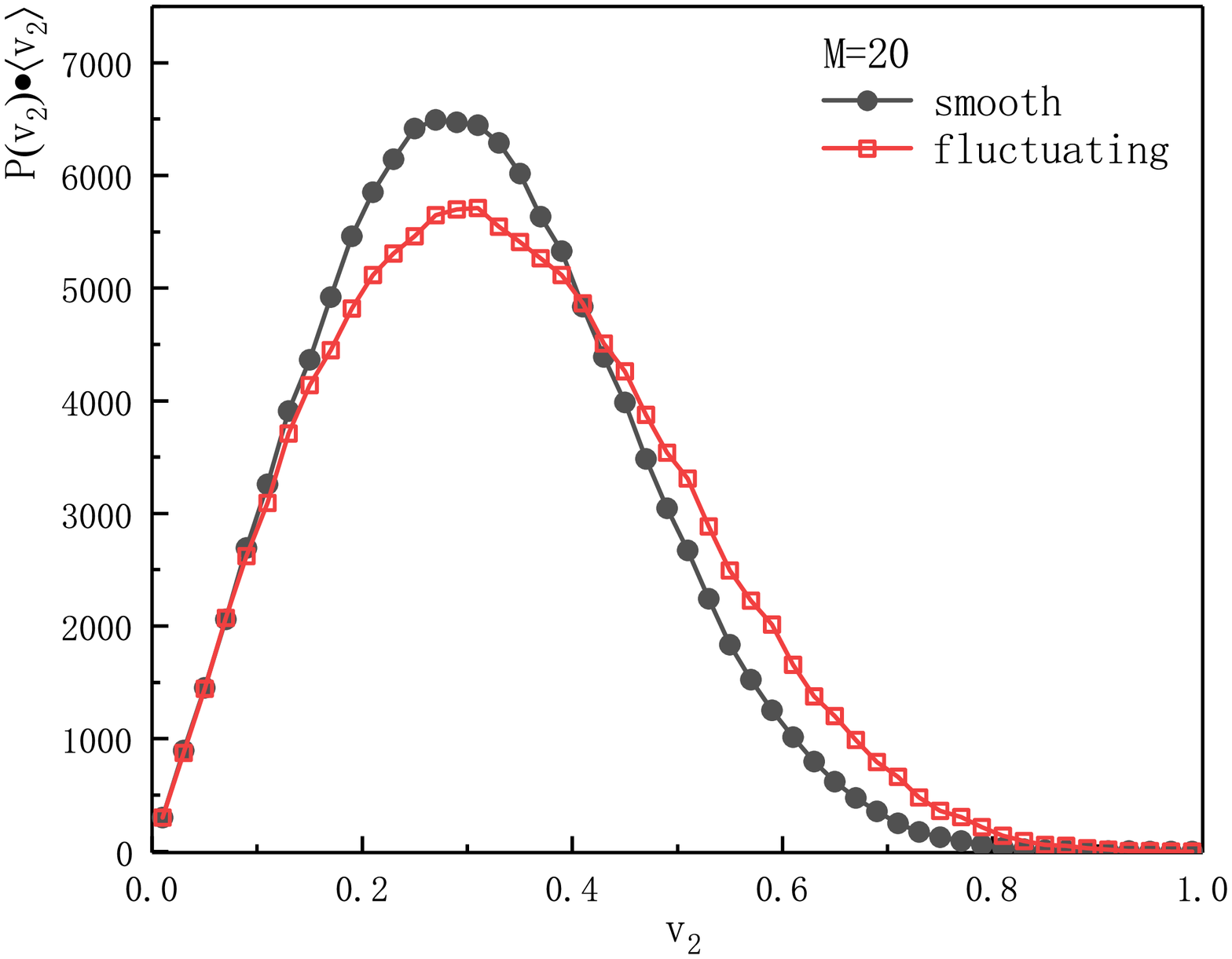}}
\end{minipage}
&
\vspace{22pt}
\begin{minipage}{250pt}
\centerline{\includegraphics[width=1.2\textwidth,height=1.0\textwidth]{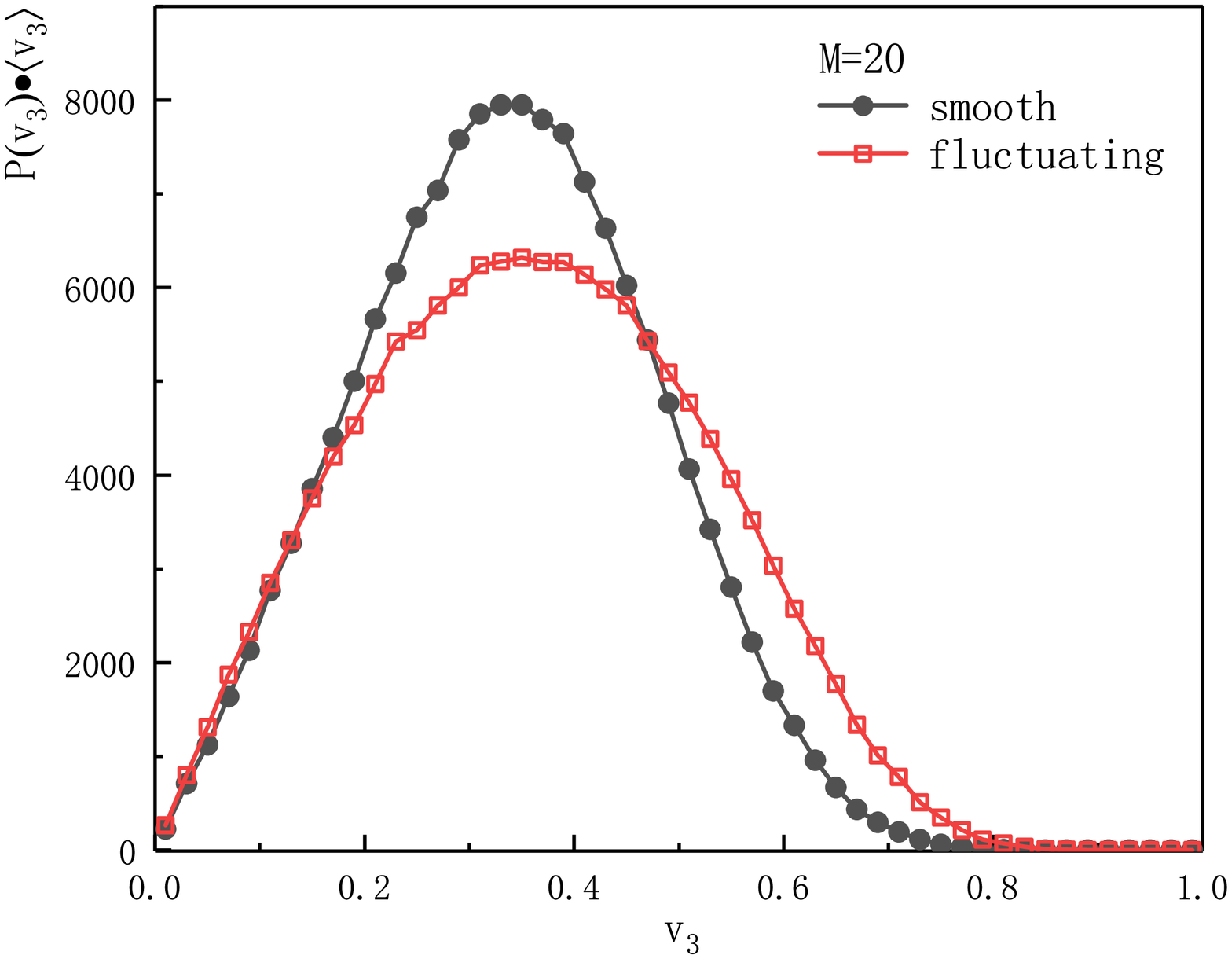}}
\end{minipage}
\\
\begin{minipage}{250pt}
\centerline{\includegraphics[width=1.2\textwidth,height=1.0\textwidth]{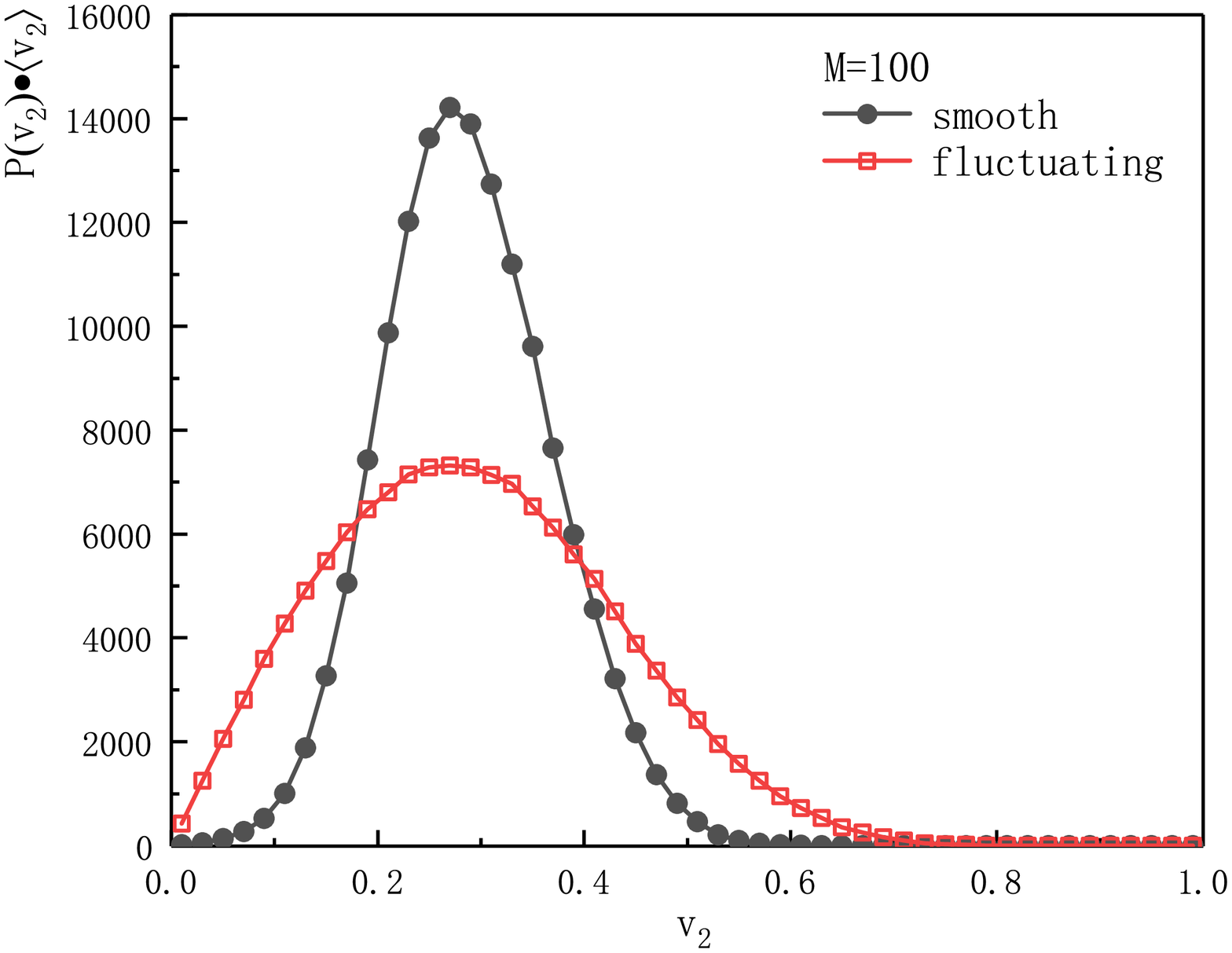}}
\end{minipage}
&
\vspace{22pt}
\begin{minipage}{250pt}
\centerline{\includegraphics[width=1.2\textwidth,height=1.0\textwidth]{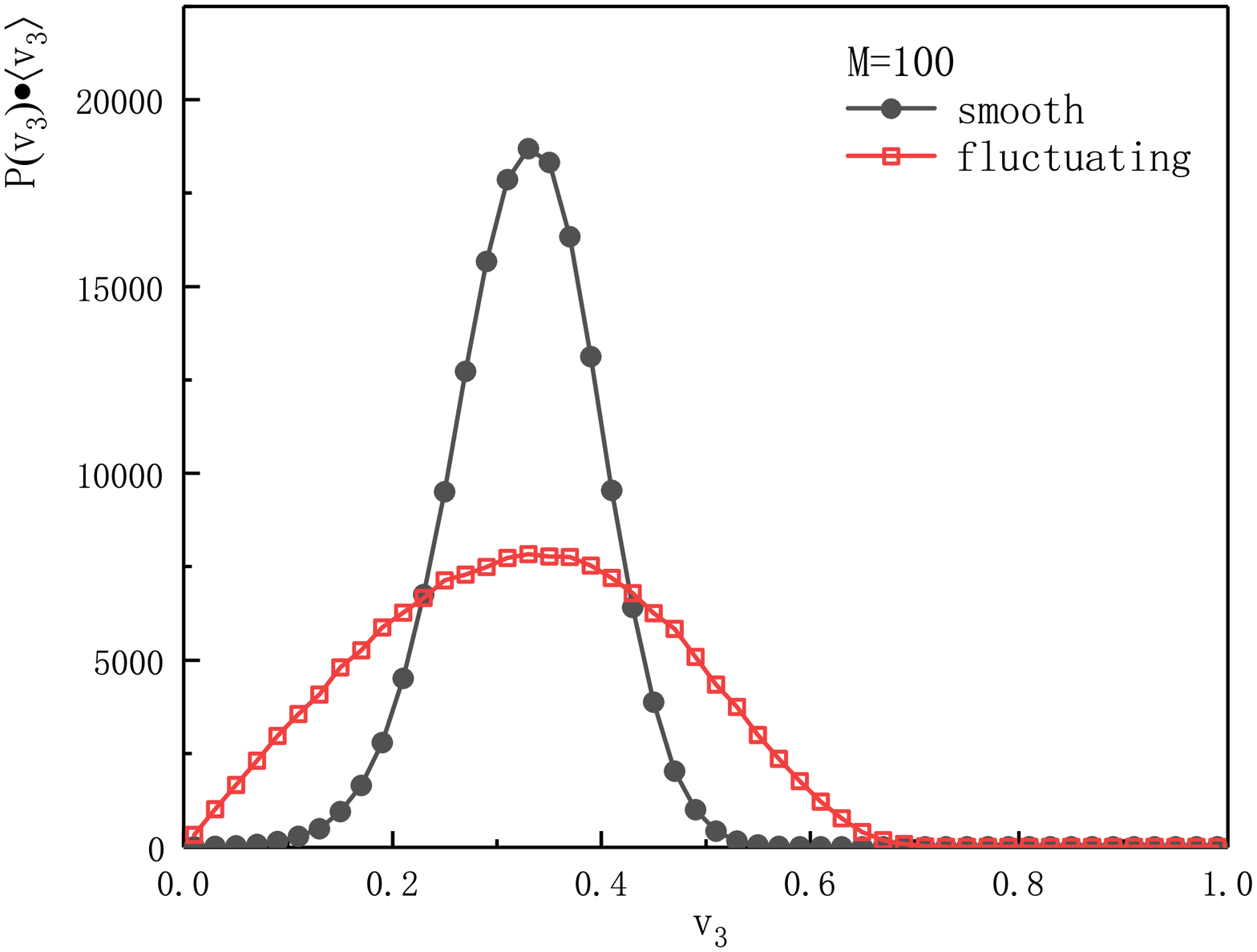}}
\end{minipage}
\end{tabular}
\renewcommand{\figurename}{Fig.}
\caption{The probability distributions of triangular flow extracted for events of smoothed (black filled dotted) and event-by-event fluctuating (red empty squares) initial conditions generated with $N=10$.
The calculations are carried out by considering different event multiplicities $M=20$ (top row) and $M=100$ (bottom row).}
\label{fig_v3_dis}
\end{figure}

In Fig.~\ref{fig_v3_dis}, we present the distribution of the elliptic and triangular flows.
Again, the distributions associated with event-by-event fluctuation initial conditions feature a more significant deviation when compared to that associated with the smooth initial conditions.
As the number of event multiplicity increases, the variance decreases owing to the suppression of statistical uncertainties, which is consistent with the results shown in Fig.~\ref{fig_v2_fluc} and Tab.~\ref{tabeveN50}.
Moreover, the difference between the fluctuating initial conditions and smooth ones becomes more significant.
Since the probability distribution of flow is a relevant observable, which can be measured experimentally, such a distinction may lead to interesting physical implications.

\begin{table}
\caption{The calculated flow harmonics and their fluctuations using the identical point-source model with $N=10$.
The calculations are carried out for smooth and fluctuating initial conditions with event multiplicity $M=10, 20, 50, 100$, and $500$.}\label{tabeveN10}
\vspace{5mm}
\begin{tabular}{cc  cccc}
        \hline\hline
          $M$          & IC & E$[v_{2}]$       & Var$[v_{2}]$        & E$[v_{3}]$ & Var$[v_{3}]$  \\
          \hline
          10            & smooth      &~~~0.375~~~&~~~$3.467\times 10^{-2}$~~~&~~~0.393~~~&~~~$3.039\times 10^{-2}$~~~\\
                            &~~~fluctuating~~~&  0.397      & $3.647\times 10^{-2}$        &  0.409      & $3.359\times 10^{-2}$   \\
          \hline
          20           & smooth      &~~~0.316~~~&~~~$2.119\times 10^{-2}$~~~&~~~0.338~~~&~~~$1.870\times 10^{-2}$~~~\\
                            &~~~fluctuating~~~&  0.341      & $2.633\times 10^{-2}$        &  0.359      & $2.480\times 10^{-2}$   \\     
          \hline                  
          50            & smooth      &  0.283      & $1.116\times 10^{-2}$        &  0.321      & $9.532\times 10^{-3}$   \\
                            & fluctuating      &  0.307      & $2.044\times 10^{-2}$        &  0.335      & $2.125\times 10^{-2}$   \\
          \hline
          100            & smooth      &  0.287      & $6.533\times 10^{-3}$        &  0.328      & $4.778\times 10^{-3}$   \\
                            & fluctuating      &  0.292      & $1.812\times 10^{-2}$        &  0.324      & $1.984\times 10^{-2}$   \\
          \hline
          500            & smooth      &  0.295      & $2.410\times 10^{-3}$        &  0.331      & $1.288\times 10^{-3}$   \\
                            & fluctuating      &  0.282      & $1.696\times 10^{-2}$        &  0.319      & $1.833\times 10^{-2}$   \\
          \hline
          1000            & smooth      &  0.297      & $2.041\times 10^{-3}$        &  0.333      & $8.467\times 10^{-4}$   \\
                            & fluctuating      &  0.289      & $1.769\times 10^{-2}$        &  0.317      & $1.832\times 10^{-2}$   \\                            
        \hline\hline
 \end{tabular}
 \end{table}
 
\begin{table}
\caption{The same as Tab.~\ref{tabeveN10}, but the calculations are carried out for $N=50$.}\label{tabeveN50}
\vspace{5mm}
\begin{tabular}{cc  cccc}
        \hline\hline
          $M$          & IC & E$[v_{2}]$       & Var$[v_{2}]$        & E$[v_{3}]$ & Var$[v_{3}]$  \\
          \hline
          10            & smooth      &~~~0.312~~~&~~~$2.332\times 10^{-2}$~~~&~~~0.342~~~&~~~$2.712\times 10^{-2}$~~~\\
                            &~~~fluctuating~~~&  0.348      & $2.879\times 10^{-2}$        &  0.347      & $2.744\times 10^{-2}$   \\
          \hline
          20           & smooth      &~~~0.249~~~&~~~$1.723\times 10^{-2}$~~~&~~~0.275~~~&~~~$1.461\times 10^{-2}$~~~\\
                            &~~~fluctuating~~~&  0.251      & $1.769\times 10^{-2}$        &  0.287      & $1.944\times 10^{-2}$   \\     
          \hline                  
          50            & smooth      &  0.193      & $7.953\times 10^{-3}$        &  0.216     & $8.454\times 10^{-3}$   \\
                            & fluctuating      &  0.215      & $1.114\times 10^{-2}$        &  0.242      & $1.341\times 10^{-2}$   \\
          \hline
          100            & smooth      &  0.173      & $4.774\times 10^{-3}$        &  0.207      & $5.051\times 10^{-3}$   \\
                            & fluctuating      &  0.186      & $8.352\times 10^{-3}$        &  0.221      & $1.136\times 10^{-2}$   \\
          \hline
          500            & smooth      &  0.167      & $9.794\times 10^{-4}$        &  0.217      & $1.069\times 10^{-3}$   \\
                            & fluctuating      &  0.169     & $6.995\times 10^{-3}$        &  0.215      & $1.070\times 10^{-2}$   \\
          \hline
          1000            & smooth      &  0.174      & $5.576\times 10^{-4}$        &  0.213      & $4.839\times 10^{-4}$   \\
                            & fluctuating      &  0.170     & $7.396\times 10^{-3}$        &  0.209      & $9.398\times 10^{-3}$   \\                            
        \hline\hline
 \end{tabular}
 \end{table}

We now proceed to analyze the above flow fluctuations from an analytic perspective.
In particular, we reassess Eqs.~\eqref{muv22} and~\eqref{sigma2v22} with the presence of initial eccentricity fluctuations.
We will restrain ourselves with the point-source model discussed in Sec.~\ref{section2} for the present study.
In this case, Eq.~\eqref{muv22} is replaced by
\begin{eqnarray}
\mathrm{E}\left[\widehat{2}_{\{2, -2\}}\right] = \mathcal{C}_\mathrm{v}^2\mathrm{E}\left[ |\epsilon_2|^2 \right] = \frac{2 \mathcal{C}_\mathrm{v}^2}{N} ,\label{muv22Mod}
\end{eqnarray}
where one assumes Eq.~\eqref{vnCen}.

Similarly, the variance Eq.~\eqref{sigma2v22} gives
\begin{eqnarray}
\mathrm{Var}\left[\widehat{2}_{\{2, -2\}}\right] 
&=& \frac{1+\mathcal{C}_\mathrm{v}^2\mathrm{E}\left[ |\epsilon_2|^2 \right] }{M(M-1)}+2\frac{M-2}{M(M-1)}\left(\mathcal{C}_\mathrm{v}^2\mathrm{E}\left[ |\epsilon_2|^2 \right] +\mathcal{C}_\mathrm{v}^3\mathrm{E}\left[ |\epsilon_2|^2|\epsilon_4| \right] \right)+ \frac{(M-2)(M-3)}{M(M-1)}\mathcal{C}_\mathrm{v}^4\mathrm{E}\left[ |\epsilon_2|^4 \right]-\mathcal{C}_\mathrm{v}^4\mathrm{E}\left[ |\epsilon_2|^2 \right]^2 \nb\\
&=&\frac{1+\frac{2 \mathcal{C}_\mathrm{v}^2}{N} }{M(M-1)}+2\frac{M-2}{M(M-1)}\left(\frac{2 \mathcal{C}_\mathrm{v}^2}{N} +\mathcal{C}_\mathrm{v}^3\mathrm{E}\left[ |\epsilon_2|^2|\epsilon_4| \right] \right)+ \frac{(M-2)(M-3)}{M(M-1)} \frac{8\mathcal{C}_\mathrm{v}^4}{2 N + N^2}-\frac{4 \mathcal{C}_\mathrm{v}^4}{N^2} . \label{sigma2v22Mod}
\end{eqnarray}
At the limit of infinite multiplicity, the above result gives 
\begin{eqnarray}
\lim\limits_{M\to \infty}\mathrm{Var}\left[\widehat{2}_{\{2, -2\}}\right] 
= \frac{8\mathcal{C}_\mathrm{v}^4}{2 N + N^2}-\frac{4 \mathcal{C}_\mathrm{v}^4}{N^2} , \label{sigma2v22lim}
\end{eqnarray}
which does not vanish as long as $N$ remains finite.
Regarding the numerical simulations, the asymptotical value Eq.~\eqref{sigma2v22lim} corresponds to the flow variance entirely due to event-by-event fluctuations.
It becomes dominant when the event multiplicity $M$ becomes rather significant so that the statistical uncertainties estimated by Eq.~\eqref{sigma2v22} vanishes.
On the other hand, flow fluctuations are completely governed by the statistical uncertainty for the events generated by smooth initial conditions.  
Compared with the results presented above in Figs.~\ref{fig_v2_fluc} and~\ref{fig_v3_dis}, the deviations of flow harmonics due to initial geometric fluctuations and statistical uncertainties are observed to be in accordance with the analytic results.

\section{Concluding remarks}\label{section5}

From a statistical viewpoint, we scrutinize the fluctuations in flow harmonics in this work.
In particular, we explore two different types of averages. 
For the first one, one assumes finite multiplicity per event but considers an infinite number of events.
For the second case, the average is taken for a given event, where one considers an infinite number of particles at the hydrodynamic limit.
Even if one assumes that the elliptic flow $v_2$ is well-defined in Eq.~\eqref{oneParDis} without any fluctuation, in the case of events with finite multiplicity, it inevitably gives rise to a finite variance.
In this context, we pointed out that one has to adopt a specific scheme to estimate flow harmonics, which is, by definition, a statistical estimator.
Due to finite statistics, it is inevitably subject to a finite variance.
From the physical perspective, it is argued that for realistic events, the flow fluctuations, demonstrated in terms of the variance of flow harmonics, carry crucial information on both initial-state geometrical and final-state statistical fluctuations.
In practice, the two types of variance will be mixed and possibly indistinguishable from that due to the initial state fluctuations, where the latter is understood to carry essential information on the underlying microscopic physical system.
A fraction of the event-by-event flow fluctuations is of pure statistical nature.
Theoretically, it will only vanish, for instance, as one generates an infinite number of particles in a hydrodynamic simulation.

In terms of experimental observables, notably collective flow and particle correlations, one of the motivations for the extensive numerical simulations for heavy-ion collisions is to distinguish the underlying microscopic mechanisms for the initial conditions.
From a hydrodynamic perspective, these distinctions may demonstrate themselves in terms of fluctuations and long-range correlations, which, in turn, might be encoded in the initial conditions by the eccentricities and their fluctuations.
In this regard, it is generally understood that hydrodynamics transform the eccentricities mostly linearly into final state flow fluctuations.
As discussed above, we argued that the fluctuations in the final flow consist of two components.
As a result, if one attempts to extract information on the fluctuations in the initial conditions via flow harmonics and their fluctuations due to finite statistics, finite multiplicity might interfere with such an effort.
Event-by-event fluctuations have long become an essential subject in the studies of relativistic heavy-ion collisions.
The present study was mainly focused on the aspect of statistical fluctuations, particularly their interplay with those originating from the finite quanta of the relevant system.
Also, as pointed out in the main text, eccentricities might emerge from the geometrical fluctuations in the initial condition where the continuum limit is taken.
Much effort has been devoted to the latter scenarios~\cite{hydro-corr-ph-07, hydro-corr-ph-08, hydro-corr-ph-11, hydro-vn-08}.
Systematically assessing or distinguishing the flow fluctuations associated with different causes is a potentially interesting topic, which might lead to further implications.
As shown in Fig.~\ref{fig_v3_dis}, the difference in flow variance between the smooth and event-by-event fluctuating initial conditions becomes more significant as the event multiplicity increases, as a result of suppression in statistical uncertainty due to large multiplicity.
This result might be physically interesting due to its observable implications.
Moreover, given experimental data, the possibility of extracting flow harmonics using a statistical estimator with less variance is also a worthy topic to explore further.

\section*{Acknowledgements}

WLQ is indebted to insightful discussions with Matthew Luzum and Jean-Yves Ollitrault.
We gratefully acknowledge the financial support from Brazilian agencies 
Funda\c{c}\~ao de Amparo \`a Pesquisa do Estado de S\~ao Paulo (FAPESP), 
Funda\c{c}\~ao de Amparo \`a Pesquisa do Estado do Rio de Janeiro (FAPERJ), 
Conselho Nacional de Desenvolvimento Cient\'{\i}fico e Tecnol\'ogico (CNPq), 
and Coordena\c{c}\~ao de Aperfei\c{c}oamento de Pessoal de N\'ivel Superior (CAPES).
This work is also supported by the National Natural Science Foundation of China.

\bibliographystyle{h-physrev}
\bibliography{references_qian}

\end{document}